\documentclass[letterpaper]{article} % DO NOT CHANGE THIS
\usepackage{aaai2026}  % DO NOT CHANGE THIS
\usepackage{times}  % DO NOT CHANGE THIS
\usepackage{helvet}  % DO NOT CHANGE THIS
\usepackage{courier}  % DO NOT CHANGE THIS
\usepackage[hyphens]{url}  % DO NOT CHANGE THIS
\usepackage{graphicx} % DO NOT CHANGE THIS
\usepackage{natbib}  % DO NOT CHANGE THIS AND DO NOT ADD ANY OPTIONS TO IT
\usepackage{caption} % DO NOT CHANGE THIS AND DO NOT ADD ANY OPTIONS TO IT
\usepackage{amsmath}
\usepackage{booktabs}
\usepackage{tikz}
\usetikzlibrary{arrows.meta,positioning,calc}
\makeatletter
\renewcommand\paragraph{\@startsection{paragraph}{4}{\parindent}{-6pt plus -2pt minus -1pt}{-1em}{\normalsize\bf}}
\makeatother

\title{Fast and Memory Efficient Multimodal Journey Planning with Delays   \thanks{Authors are listed in alphabetical order.}}
\author{
    Denys Katkalo\textsuperscript{1},
    Andrii Rohovyi\textsuperscript{2},
    Toby Walsh\textsuperscript{2}
}
\affiliations{
    \textsuperscript{1}Igor Sikorsky Kyiv Polytechnic Institute, Kyiv, Ukraine \\
    \textsuperscript{2} School of Computer Science and Engineering, University of New South Wales, Sydney, Australia \\
    katkalo.denys@lll.kpi.ua, \{a.rohovyi, t.walsh\}@unsw.edu.au
}
\nocopyright
\begin{document}

\maketitle

\begin{abstract}

State-of-the-art multimodal journey-planning algorithms, such as ULTRA, have recently been adapted to account for delays. In this work, we extend this approach to be more memory-efficient, faster, and accurate. We also adapt this framework to other state-of-the-art algorithms, like CSA and RAPTOR. We demonstrate a speedup of $1.9\text{--}4.2\times$ over existing algorithms in the single-objective search (earliest arrival time). In the bicriteria setting, we achieve competitive speedup results but greater accuracy. We also find that our method scales much better as the delay buffer $\Delta$ increases.

\end{abstract}

\section{Introduction}
Multimodal public transit journey planning seeks the earliest arrival time at a target stop (\emph{single-objective}) or the Pareto set over arrival time and number of trips (\emph{bicriteria}), under a transit timetable combined with unrestricted walking transfers.
Real timetables admit delays bounded above by an input parameter $\Delta$, the \emph{delay buffer}; the challenge is to answer such queries correctly under any delay scenario in $[0, \Delta]$.

In multimodal journey planning, where there are transfers between different modes of transport, the state-of-the-art framework ULTRA~\cite{ultra_mcraptor,baum2019ultra,ultra} (\emph{UnLimited TRAnsfers}) precomputes shortcut transfers that allow fast queries with unlimited transfer distances, but this technique involves heavy precomputation.
Its delay-aware extension, Delay-ULTRA (D-ULTRA)~\cite{bez2024delay,sauer2024closing}, annotates each shortcut with a delay interval so that shortcuts can be activated or deactivated as delays are revealed.
However, D-ULTRA operates exclusively at the level of individual vehicle events and targets only Trip-Based (TB) routing~\cite{trip_based_routing}.

In this work, we project D-ULTRA's event-level shortcuts onto the stops they connect, producing a much smaller stop-level shortcut set.
This transformation makes ULTRA-CSA~\cite{csa} (\emph{Connection Scan Algorithm}) and ULTRA-RAPTOR~\cite{raptor} (round-based router) usable in the delay setting alongside D-ULTRA-TB, and yields three improvements.

\paragraph{Memory efficiency.}
The stop-level projection compresses the shortcut set by $48\text{--}180\times$ in edge count and $141\text{--}375\times$ in memory (from gigabytes to megabytes), because many event-level shortcuts between distinct events at the same pair of stops collapse into a single edge.

\paragraph{Speed.}
CSA with stop-level shortcuts (D-ULTRA-CSA) achieves speedups of $1.9\text{--}4.2\times$ over D-ULTRA-TB in the single-objective setting and up to $12.2\times$ over the MR baseline (Multimodal RAPTOR)~\cite{delling2013computing}.
In the bicriteria setting, RAPTOR with stop-level shortcuts is competitive with D-ULTRA-TB on both networks tested.

\paragraph{Accuracy.}
D-ULTRA-TB misses optimal journeys on both networks at every delay buffer tested.
With stop-level shortcuts and replacement, our algorithms produce zero errors on all configurations.

Moreover, stop-level shortcuts scale well with the delay buffer: increasing $\Delta$ grows the event-level set by about $14.7\times$ and slows D-ULTRA-TB by $2.3\times$, while the stop-level set grows only $4.6\times$ and query times increase by at most 20\%.

\subsection{Related Work}

\subsubsection{Transit routing algorithms.}
Public transit routing has evolved from Dijkstra-based algorithms to specialised approaches such as RAPTOR~\cite{raptor}, CSA~\cite{csa}, TB-routing~\cite{trip_based_routing, trip_based_routing_condense}, and Transfer Patterns~\cite{transfer_patterns}. Many of these limit transfer distance, hampering integration of secondary modes such as e-scooters, bikes, or taxis. For multimodal pathfinding, algorithms fall into two categories: those without precomputation, which can work directly on delayed timetables, and those that require precomputation.

For exact delay-aware routing without shortcut precomputation, MR~\cite{delling2013computing} computes Pareto-optimal journeys by combining Dijkstra~\cite{dijkstra1959note} on the transfer graph with RAPTOR's round structure. It can be applied directly to a delayed timetable but is relatively slow. An alternative is the Hub Labeling (HL) version of RAPTOR~\cite{phan2019hl,abraham2011hub,delling2016hub}, which is often omitted from comparison analyses~\cite{bez2024delay} due to its marginal improvement over MR, but it still produces faster bicriteria responses than MR~\cite{sauer2024closing}. There is also a single-objective CSA adaptation that works with HL, but to our knowledge it has not previously been compared against MR on the same machine and query set. We address this gap.

A faster single-objective alternative that requires no preprocessing is classical Time-Dependent Dijkstra (TD-Dijkstra)~\cite{rohovyi2025multimodal, rohovyi2026tad}, which requires dominated-connection filtering that can produce incorrect results on networks with buffer times. Transfer Aware Dijkstra (TAD)~\cite{rohovyi2026tad} addresses this limitation by replacing per-edge relaxation with trip-level scanning, providing exact answers on networks with or without buffer times. MR, HL-RAPTOR, HL-CSA, and TAD serve as ground-truth baselines in our experiments, since they work directly on the delayed timetable. TD-Dijkstra is also evaluated but is not applicable to networks with buffer times.

Beyond multimodal transit, generic temporal-planning approaches handle delays in transit settings (e.g., CSA~\cite{csa} on a time-expanded timetable representation and the dynamic-replanning formulation of Abuaisha et al.~\cite{abuaisha2025dynamic}), but both restrict walking distance by assuming a transitively closed transfer graph. Wagner and Z\"{u}ndorf~\cite{unrestricted_walking} showed this restriction is not benign: on the German network, median nighttime travel times differ by up to two hours between restricted and unrestricted variants, and 75\% of daytime queries return suboptimal paths. Lifting it while keeping query times competitive motivates the ULTRA family.

\subsubsection{Transfer-graph acceleration and shortcut precomputation.}
We use hierarchical transfer-graph structures, namely Contraction Hierarchies (CH)~\cite{geisberger2012ch}, Core-CH~\cite{bauer2010combining,baum2019ultra}, Bucket-CH~\cite{knopp2007manytoMany}, and Hub Labeling (HL)~\cite{abraham2011hub,delling2016hub}; full definitions appear in the Preliminaries. For shortcut precomputation in multimodal pathfinding, ULTRA~\cite{ultra, ultra_mcraptor} is the leading solution. Bez's bachelor thesis~\cite{bez2020thesis} initiated the delay-tolerant line, introducing DB-ULTRA: \emph{unannotated stop-level shortcuts} valid for any delay scenario within a fixed buffer~$\Delta$, evaluated with ULTRA-RAPTOR and ULTRA-CSA on Switzerland and Germany. The thesis observed rapid growth in shortcut count under the Delay-All model (over twelve million at $\Delta = 30$\,min on Switzerland) and proposed two directions to address it: annotating shortcuts with delay sub-intervals, and adapting the framework to Trip-Based Routing~\cite{trip_based_routing}. Bez and Sauer~\cite{bez2024delay,sauer2024closing} subsequently combined both into event-level Trip-Based shortcuts with delay annotations, the algorithm we benchmark against. We propose a third direction: rather than re-deriving shortcuts natively at the stop level, we project the event-level shortcut set onto the stops, dropping annotations in the process, combining the coverage of the event-level construction with the compactness the annotations were designed to provide.

\subsubsection{Algorithms compared in this work.}
MR~\cite{delling2013computing}, TAD~\cite{rohovyi2026tad}, HL-RAPTOR, and HL-CSA produce ground-truth results, since they do not require shortcut precomputation and can be run directly on the delayed timetable. TD-Dijkstra is faster but can produce incorrect results on networks with buffer times, such as Switzerland in our experiments.

D-ULTRA-CSA, HL-CSA, TAD, and TD-Dijkstra are single-objective algorithms, returning the path with the earliest arrival time.
MR, HL-RAPTOR, D-ULTRA-TB, D-ULTRA-RAPTOR, and D-ULTRA-RAPTOR~(EP) return bicriteria Pareto-optimal results, optimising arrival time and number of trips.

D-ULTRA-CSA, D-ULTRA-RAPTOR, and D-ULTRA-RAPTOR~(EP) use the stop-level delay shortcuts as their transfer graph.
D-ULTRA-TB runs Trip-Based routing with the event-level D-ULTRA shortcuts of~\cite{bez2024delay}, using their delay annotations to decide which shortcuts to activate during the query.
HL-RAPTOR and HL-CSA~\cite{phan2019hl,abraham2011hub,delling2016hub,ultra} replace Core-CH with hub labels computed on the transfer graph (preprocessing times in Table~\ref{tab:instances}).
Because hub labels encode shortest-path distances in the schedule-independent transfer graph, HL algorithms require no shortcut precomputation and are inherently delay-robust.

\section{Preliminaries}
This section defines key terminology and presents the foundational algorithms.

\subsection{Terminology}
We introduce the data model used throughout the paper.

\paragraph{Network.}
A public transit network consists of a set of stops~$\mathcal{S}$, a set of trips~$\mathcal{T}$, a set of routes, and a transfer graph~$G = (V, E)$. Stops are physical locations: bus platforms, train stations, ferry terminals, where passengers board or alight vehicles.

A trip $T \in \mathcal{T}$ models a single vehicle run as a sequence of stop events $\langle \varepsilon_0, \ldots, \varepsilon_k \rangle$. Each stop event $\varepsilon$ records the stop it serves, written $v(\varepsilon) \in \mathcal{S}$, together with an arrival time $\tau_{\text{arr}}(\varepsilon)$ and a departure time $\tau_{\text{dep}}(\varepsilon)$. A route groups trips that visit the same sequence of stops; trips within a route differ only in their scheduled times.

The transfer graph $G = (V, E)$ captures how passengers move between stops on foot or by other non-scheduled modes. Its vertex set satisfies $\mathcal{S} \subseteq V$ and may contain additional intermediate vertices representing street-network intersections or points of interest. Every edge $e = (u, w) \in E$ carries a travel time $\tau_{\text{tra}}(e)$. We impose no restrictions on~$G$: it need not be transitively closed, it may contain cycles, and travel times may reflect walking, cycling, or any other schedule-independent mode.

Algorithms that process transfers in a single relaxation step per round, notably CSA and RAPTOR, require the transfer graph to be transitively closed among the stops it connects. A graph is transitively closed if for every pair of edges $(a, b), (b, c) \in E$ the edge $(a, c)$ also belongs to~$E$ with $\tau_{\text{tra}}((a, c)) \leq \tau_{\text{tra}}((a, b)) + \tau_{\text{tra}}((b, c))$. Computing this closure is the primary bottleneck that limits transfer distances in practice~\cite{unrestricted_walking}, and overcoming it is the central motivation for shortcut-based approaches such as ULTRA.

\paragraph{Shortcuts.}
ULTRA precomputes shortcut edges that allow query algorithms to find optimal journeys without exploring the full transfer graph. Shortcuts can be indexed at two granularities.

An \emph{event-level shortcut} is an edge $(\varepsilon_a, \varepsilon_b)$ connecting two stop events from different trips, together with a transfer travel time $\tau_{\text{tra}}(\varepsilon_a, \varepsilon_b)$. It encodes that a walking path of this duration from $v(\varepsilon_a)$ to $v(\varepsilon_b)$ may be part of an optimal journey. We denote the full set of event-level shortcuts by $\mathcal{E}_{\text{sc}}$. D-ULTRA~\cite{bez2024delay} further annotates each event-level shortcut with a delay interval $[\delta_{\min}, \delta_{\max}]$ specifying under which delay realisations the shortcut remains necessary. Each event-level shortcut thus stores four attributes: destination event, travel time, minimum delay, and maximum delay. The full event-level graph is stored as a standalone adjacency array indexed by stop events, which also carries per-vertex data for all $|\mathcal{E}|$ stop-event vertices.

A \emph{stop-level shortcut} is an edge $(s, t)$ connecting two stops $s, t \in \mathcal{S}$, storing only the destination stop and the travel time. Since stop-level shortcuts carry no delay annotations, they are always active regardless of the delay scenario. The stop-level shortcut graph $G_{\text{sc}} = (\mathcal{S}, E_{\text{sc}})$ is obtained by projecting the event-level shortcuts onto the stops they serve; we define the projection precisely in Section~\ref{sec:stop_to_stop}.

Event-level shortcuts are a natural fit for TB-routing, which indexes transfers by stop event. Stop-level shortcuts are directly compatible with CSA and RAPTOR, which index transfers by stop and treat shortcuts as ordinary edges in the transfer graph.

\subsection{Algorithms}

\paragraph{Dijkstra and TD-Dijkstra.}
Dijkstra's algorithm~\cite{dijkstra1959note} computes shortest paths from a single source by maintaining a priority queue of tentative distances and repeatedly settling the nearest unsettled vertex.
In a time-dependent network, edge travel times depend on when the edge is traversed.
Time-Dependent Dijkstra (TD-Dijkstra)~\cite{multi_dijkstra} adapts the classical algorithm by evaluating each edge weight at the current arrival time of its tail vertex, preserving optimality provided the network satisfies the FIFO property (earlier departure implies earlier arrival).
Public transit schedules can violate FIFO when express and local services share stops; efficient implementations restore FIFO by filtering dominated connections during preprocessing.

\paragraph{Transfer Aware Dijkstra (TAD).}
Many transit networks specify a buffer time at each stop: a minimum waiting time that transferring passengers must respect before boarding a connecting service.
The dominated-connection filtering used by TD-Dijkstra is unsound when buffer times are present, because it cannot distinguish seated-through passengers from transferring ones~\cite{rohovyi2026tad}.
TAD~\cite{rohovyi2026tad} addresses this by replacing per-edge relaxation with trip-level scanning, preserving correctness on networks with arbitrary buffer times. Two variants have been tested: with Core-CH and with Bucket-CH. In our experiments TAD uses Bucket-CH as the faster option.

We additionally evaluate a classical TD-Dijkstra variant that applies dominated-connection filtering and uses Bucket-CH for acceleration.
This variant is faster than TAD but is not applicable to networks with buffer times.

\paragraph{CSA.}
The Connection Scan Algorithm~\cite{csa} takes an array of all elementary connections sorted by departure time and performs a single linear sweep.
For each connection whose departure stop has already been reached, the algorithm updates the arrival time at the connection's arrival stop.

\paragraph{RAPTOR.}
RAPTOR~\cite{raptor} organises the search into rounds, where round~$k$ finds the best journeys using at most~$k$ vehicle trips.
Within each round the algorithm scans every route that serves a stop whose arrival time improved in the previous round, propagates arrival times along the route's stop sequence, and then relaxes all outgoing transfer edges from every improved stop. This round structure naturally produces a Pareto set over arrival time and number of trips.

\paragraph{Early Pruning (EP).}
Early Pruning~\cite{rohovyi2026ep} is a technique that accelerates the transfer relaxation phase of RAPTOR and its variants without affecting optimality. The key observation is that if the outgoing transfer edges at each stop are pre-sorted by travel time, the algorithm can terminate the transfer loop as soon as the next transfer would produce an arrival time later than the current best arrival at the target.

\paragraph{Contraction Hierarchies (CH).}
CH~\cite{geisberger2012ch} is a preprocessing technique for accelerating shortest-path queries in static graphs.
Vertices are contracted in a heuristically determined order: removing a vertex while inserting shortcut edges between its neighbours to preserve shortest-path distances.
The result is an augmented graph that decomposes into an upward graph (edges from lower-ranked to higher-ranked vertices) and a downward graph (the reverse).
Queries are answered by bidirectional Dijkstra, with the forward search exploring the upward graph and the backward search exploring the downward graph.

\paragraph{Core-CH.}
Core-CH~\cite{delling2013computing,baum2019ultra} adapts CH to multimodal networks by leaving a set of core vertices uncontracted, with all stops included in the core ($\mathcal{S} \subseteq V_c$).
This produces a core graph over which Dijkstra searches are run during MR's transfer relaxation phase.
To prevent the core graph from growing too large, contraction is stopped once the average vertex degree in the core exceeds a specified threshold.

\paragraph{Bucket-CH.}
Bucket-CH~\cite{knopp2007manytoMany,geisberger2012ch} extends CH to handle one-to-many queries efficiently.
After standard CH preprocessing, a backward search from each target vertex populates a bucket at every settled vertex with the distance to that target.
A subsequent forward search from the source then scans these buckets to compute all source-to-target distances in a single pass.
In our setting, TAD and TD-Dijkstra use Bucket-CH for initial-transfer acceleration, and D-ULTRA's replacement search uses it to find new event-level shortcuts during the update phase.

\paragraph{MR.}
MR~\cite{delling2013computing} is a modification of RAPTOR for the unlimited transfer problem. Instead of requiring a transitively closed transfer graph, MR runs Dijkstra searches accelerated by Core-CH during the transfer relaxation phase of each round.

\paragraph{Hub Labeling.}
An alternative to Core-CH for accelerating transfer relaxation is Hub Labeling (HL)~\cite{abraham2011hub,delling2016hub}. Phan and Viennot~\cite{phan2019hl} applied HL to both RAPTOR and CSA for unrestricted walking; HL-RAPTOR was shown to be marginally faster than MR in~\cite{ultra}, but HL-CSA has not been evaluated in the delay setting. We include both to provide delay-robust baselines that require no shortcut precomputation.

\paragraph{Trip-Based Routing.}
Trip-Based (TB) routing~\cite{trip_based_routing} replaces route scanning with a direct search over trips.
It maintains, for each trip, the earliest stop at which the trip can be boarded, and expands from one trip to the next via a precomputed transfer graph linking stop events.
Because transfers are resolved at the event level, TB can exploit tighter pruning: a transfer from event~$\varepsilon_a$ to event~$\varepsilon_b$ is only used if the traveller arrives at~$v(\varepsilon_a)$ before $\tau_{\text{dep}}(\varepsilon_b)$ minus the walking time.
This event-level granularity is the reason D-ULTRA~\cite{bez2024delay} targets TB: delay intervals can be attached to individual event-level shortcuts.
However, the same granularity also makes the shortcut set vulnerable to coverage gaps when delays shift trip timings, as we demonstrate experimentally.

\paragraph{ULTRA.}
UnLimited TRAnsfers (ULTRA)~\cite{ultra} decouples the transfer precomputation from the query algorithm.
During preprocessing, a witness--candidate Dijkstra search identifies, for each pair of consecutive vehicle trips in an optimal journey, the shortest transfer path through the full street network.
These paths are stored as shortcut edges in the transfer graph, after which any query algorithm: CSA, RAPTOR, or TB can use the augmented graph as a drop-in replacement for the original transfer graph.
The key insight is that the number of necessary shortcuts is far smaller than the full transitive closure, because most stop pairs are never connected by an optimal two-trip journey.
ULTRA has been extended to multicriteria search~\cite{ultra_mcraptor}.
D-ULTRA~\cite{bez2024delay,sauer2024closing} further extends the precomputation to account for bounded delays, producing event-level shortcuts annotated with delay intervals that indicate under which delay scenarios each shortcut remains feasible.

When a delay scenario is applied at query time, D-ULTRA's update phase adjusts the shortcut set in one of two modes~\cite{bez2024delay}.
The \emph{basic} update removes shortcuts that have become infeasible under the current delays: either the transfer is no longer physically possible (the passenger arrives too late to board), or the realised delay falls outside the shortcut's annotated interval.
The \emph{advanced} update additionally performs a heuristic replacement search: for each origin event whose shortcuts were invalidated, a Bucket-CH search finds new event-level shortcuts that are feasible under the current delay scenario, partially recovering the transfer coverage lost by the removal step.
The number of replacement shortcuts is small relative to the full set (typically below $0.2\%$), but as we show experimentally, even this modest recovery noticeably reduces D-ULTRA-TB's error rate.

\section{Stop-Level Delay Shortcuts}\label{sec:stop_to_stop}

D-ULTRA~\cite{bez2024delay,sauer2024closing} computes shortcuts at the event level: each shortcut $(\varepsilon_a, \varepsilon_b) \in \mathcal{E}_{\text{sc}}$ connects two specific stop events and carries a delay interval indicating under which delay realisations the shortcut is necessary. This granularity is a natural fit for TB-routing, which indexes transfers by stop event, but it is incompatible with CSA and RAPTOR, which index transfers by stop. We propose a stop-level alternative that projects event-level shortcuts onto the stops they serve.

\subsection{Delay Model}

Each stop event~$\varepsilon$ may experience an arrival delay $\delta_{\text{arr}}(\varepsilon) \in [0, \Delta]$ and a departure delay $\delta_{\text{dep}}(\varepsilon) \in [0, \Delta]$, yielding delayed times
\begin{align}
  \tilde{\tau}_{\text{arr}}(\varepsilon) &= \tau_{\text{arr}}(\varepsilon) + \delta_{\text{arr}}(\varepsilon), \label{eq:delay_arr}\\
  \tilde{\tau}_{\text{dep}}(\varepsilon) &= \tau_{\text{dep}}(\varepsilon) + \delta_{\text{dep}}(\varepsilon). \label{eq:delay_dep}
\end{align}
A \emph{delay scenario} is a function $\delta$ assigning a delay pair $(\delta_{\text{arr}}(\varepsilon), \delta_{\text{dep}}(\varepsilon))$ to every stop event~$\varepsilon$. We write $\delta = 0$ for the \emph{best-case} scenario where all delays are zero and $\delta = \Delta$ for the \emph{worst-case} scenario where every delay takes its maximum value. The set of all valid delay scenarios is $\mathcal{D} = \{\delta : \forall\, \varepsilon,\; 0 \le \delta_{\text{arr}}(\varepsilon) \le \Delta \;\wedge\; 0 \le \delta_{\text{dep}}(\varepsilon) \le \Delta\}$.

\subsection{Projection from Event Level to Stop Level}

Given the set of event-level shortcuts $\mathcal{E}_{\text{sc}}$ produced by D-ULTRA, we construct the stop-level shortcut graph $G_{\text{sc}} = (\mathcal{S}, E_{\text{sc}})$ as follows. For each pair of stops $(s, t)$ connected by at least one event-level shortcut, we insert a single stop-level edge:
\begin{equation}\label{eq:stop_edges}
\begin{gathered}
E_{\text{sc}} = \bigl\{(s, t) \in \mathcal{S} \times \mathcal{S} : \exists\, (\varepsilon_a, \varepsilon_b) \in \mathcal{E}_{\text{sc}} \\
\text{with}\; v(\varepsilon_a) = s \;\text{and}\; v(\varepsilon_b) = t \bigr\}.
\end{gathered}
\end{equation}
Each such edge carries the minimum travel time over all contributing event-level shortcuts:
\begin{equation}\label{eq:stop_time}
\tau_{\text{tra}}(s, t) = \min_{\substack{(\varepsilon_a, \varepsilon_b) \in \mathcal{E}_{\text{sc}} \\ v(\varepsilon_a) = s,\; v(\varepsilon_b) = t}} \tau_{\text{tra}}(\varepsilon_a, \varepsilon_b).
\end{equation}
The stop-level graph $G_{\text{sc}}$ is merged into the existing transfer graph $G$ as a set of new edges between stop vertices, requiring no additional vertex allocation. Event-level and stop-level shortcuts are stored in two separate data structures; for CSA and RAPTOR, the stop-level graph fully replaces the event-level set rather than augmenting it as a separate layer.

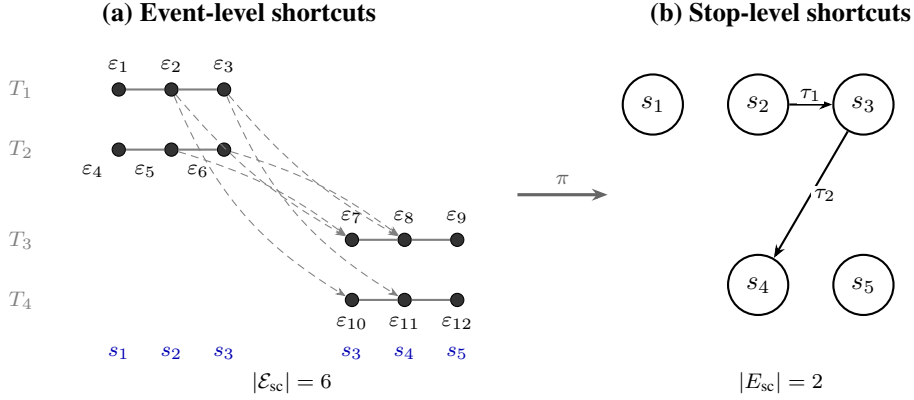
\begin{figure*}[t]
\centering
\begin{tikzpicture}[
  event/.style={circle, draw, fill=black!80, inner sep=0pt, minimum size=5pt},
  stop/.style={circle, draw, thick, fill=white, inner sep=0pt, minimum size=8mm, font=\small},
  tripline/.style={thick, gray},
  esc/.style={-{Stealth[length=4pt]}, thin, densely dashed, black!50},
  ssc/.style={-{Stealth[length=5pt]}, thick, black},
  lbl/.style={font=\scriptsize, fill=white, inner sep=1pt},
]

% ── Left panel: event-level ──
\node[font=\small\bfseries] at (2.3,4.0) {(a) Event-level shortcuts};

% Trip T1
\node[font=\scriptsize, gray] at (-0.6,3.0) {$T_1$};
\foreach \i/\lab/\slab in {1/\varepsilon_1/s, 2/\varepsilon_2/s, 3/\varepsilon_3/s} {
  \node[event, label={[font=\scriptsize]above:$\lab$}] (e1\i) at ({0.7*\i},3.0) {};
}
\draw[tripline] (e11) -- (e12) -- (e13);

% Trip T2
\node[font=\scriptsize, gray] at (-0.6,2.2) {$T_2$};
\foreach \i/\lab in {1/\varepsilon_4, 2/\varepsilon_5, 3/\varepsilon_6} {
  \node[event, label={[font=\scriptsize]below left:$\lab$}] (e2\i) at ({0.7*\i},2.2) {};
}
\draw[tripline] (e21) -- (e22) -- (e23);

% Trip T3
\node[font=\scriptsize, gray] at (-0.6,1.0) {$T_3$};
\foreach \i/\lab in {1/\varepsilon_7, 2/\varepsilon_8, 3/\varepsilon_9} {
  \node[event, label={[font=\scriptsize]above:$\lab$}] (e3\i) at ({3.1+0.7*\i},1.0) {};
}
\draw[tripline] (e31) -- (e32) -- (e33);

% Trip T4
\node[font=\scriptsize, gray] at (-0.6,0.2) {$T_4$};
\foreach \i/\lab in {1/\varepsilon_{10}, 2/\varepsilon_{11}, 3/\varepsilon_{12}} {
  \node[event, label={[font=\scriptsize]below:$\lab$}] (e4\i) at ({3.1+0.7*\i},0.2) {};
}
\draw[tripline] (e41) -- (e42) -- (e43);

% Stop labels under event columns
\node[font=\scriptsize, blue!70!black] at (0.7,-0.5) {$s_1$};
\node[font=\scriptsize, blue!70!black] at (1.4,-0.5) {$s_2$};
\node[font=\scriptsize, blue!70!black] at (2.1,-0.5) {$s_3$};
\node[font=\scriptsize, blue!70!black] at (3.8,-0.5) {$s_3$};
\node[font=\scriptsize, blue!70!black] at (4.5,-0.5) {$s_4$};
\node[font=\scriptsize, blue!70!black] at (5.2,-0.5) {$s_5$};

% Event-level shortcuts (3 for s2→s3, 3 for s3→s4)
\draw[esc] (e12) to[bend right=10] (e31);
\draw[esc] (e12) to[bend right=18] (e41);
\draw[esc] (e22) to[bend left=10] (e31);
\draw[esc] (e13) to[bend right=10] (e32);
\draw[esc] (e13) to[bend right=18] (e42);
\draw[esc] (e23) to[bend left=10] (e32);

% Count labels at bottom
\node[font=\scriptsize] at (3.0,-0.9) {$|\mathcal{E}_{\text{sc}}| = 6$};

% ── Arrow between panels ──
\draw[-{Stealth[length=6pt]}, very thick, black!60] (6.0,1.6) -- node[above, font=\scriptsize] {$\pi$} (7.2,1.6);

% ── Right panel: stop-level ──
\node[font=\small\bfseries] at (9.5,4.0) {(b) Stop-level shortcuts};

\node[stop] (S1) at (7.8,2.8) {$s_1$};
\node[stop] (S2) at (9.2,2.8) {$s_2$};
\node[stop] (S3) at (10.6,2.8) {$s_3$};
\node[stop] (S4) at (9.2,0.4) {$s_4$};
\node[stop] (S5) at (10.6,0.4) {$s_5$};

% Stop-level shortcuts
\draw[ssc] (S2) -- node[lbl, above] {$\tau_1$} (S3);
\draw[ssc] (S3) -- node[lbl, right] {$\tau_2$} (S4);

% Count label
\node[font=\scriptsize] at (9.5,-0.9) {$|E_{\text{sc}}| = 2$};

\end{tikzpicture}
\caption{Projection $\pi$ from event-level to stop-level shortcuts. (a)~Trips $T_1$, $T_2$ serve stops $s_1, s_2, s_3$ and trips $T_3$, $T_4$ serve stops $s_3, s_4, s_5$. Six event-level shortcuts (dashed arrows) connect individual stop events across the two trip groups. (b)~The projection merges all shortcuts between the same pair of stops into a single edge with the minimum travel time, yielding two stop-level shortcuts. The travel times are $\tau_1 = \min\{\tau_{\text{tra}}(\varepsilon_2, \varepsilon_7),\, \tau_{\text{tra}}(\varepsilon_2, \varepsilon_{10}),\, \tau_{\text{tra}}(\varepsilon_5, \varepsilon_7)\}$ and $\tau_2 = \min\{\tau_{\text{tra}}(\varepsilon_3, \varepsilon_8),\, \tau_{\text{tra}}(\varepsilon_3, \varepsilon_{11}),\, \tau_{\text{tra}}(\varepsilon_6, \varepsilon_8)\}$.}\label{fig:projection}
\end{figure*}

Since many event-level shortcuts between distinct events at the same pair of stops collapse into a single stop-level edge, $|E_{\text{sc}}| \ll |\mathcal{E}_{\text{sc}}|$ in practice (see Table~\ref{tab:instances}). Figure~\ref{fig:projection} illustrates this construction.

The projection can equivalently be expressed as a map $\pi \colon \mathcal{E}_{\text{sc}} \to E_{\text{sc}}$ defined by
\begin{equation}\label{eq:projection}
\pi(\varepsilon_a, \varepsilon_b) = \bigl(v(\varepsilon_a),\; v(\varepsilon_b)\bigr),
\end{equation}
which sends every event-level shortcut to the stop pair it connects. The stop-level edge set is the image $E_{\text{sc}} = \pi(\mathcal{E}_{\text{sc}})$, and the travel time on each stop-level edge is the minimum over its preimage (Equation~\eqref{eq:stop_time}).
Since many trips visit the same stops, the preimage of a single stop pair $(s, t)$ can contain hundreds of event-level shortcuts, all of which collapse into one edge. This explains the count compression ratios observed in Table~\ref{tab:instances}: $48\text{--}180\times$ across all configurations; memory compression is even higher ($141\text{--}375\times$) because the projection also discards the per-edge delay annotations.

\subsection{Accuracy}

The stop-level shortcut set $E_{\text{sc}}$ is a conservative superset of the event-level set at the stop-pair level. To see why, consider a transfer from stop~$s$ to stop~$t$ and suppose the event-level set contains at least one shortcut $(\varepsilon_a, \varepsilon_b) \in \mathcal{E}_{\text{sc}}$ with $v(\varepsilon_a) = s$ and $v(\varepsilon_b) = t$. By construction, $\pi(\varepsilon_a, \varepsilon_b) = (s, t) \in E_{\text{sc}}$, so the stop-level graph contains this edge. Moreover, $\tau_{\text{tra}}(s, t) \le \tau_{\text{tra}}(\varepsilon_a, \varepsilon_b)$ by the minimum in Equation~\eqref{eq:stop_time}, so the stop-level travel time is at most as large. We can therefore state the following property.

\paragraph{Superset Property.}
For every pair of stops $(s, t)$ covered by at least one event-level shortcut in $\mathcal{E}_{\text{sc}}$, the edge $(s, t)$ is present in~$G_{\text{sc}}$ with a travel time no greater than that of any contributing event-level shortcut.
Since D-ULTRA's precomputation is designed to produce event-level shortcuts for all transfers needed under any delay scenario $\delta \in \mathcal{D}$, the stop-level graph inherits this coverage at the stop-pair level.

The converse does not hold: $E_{\text{sc}}$ may contain edges that are not needed under a particular delay scenario. An unnecessary shortcut may cause the query algorithm to attempt boarding a connection that is in fact infeasible (because the passenger arrives too late), but the algorithm will simply fail to board and fall back to alternative paths. This conservatism therefore does not harm accuracy; it can only introduce a marginal amount of extra work during the query.

In contrast, D-ULTRA~\cite{bez2024delay} keeps shortcuts event-specific: each shortcut applies only to a particular pair of stop events. When delays shift trip timings, the event-level set may lack a precomputed shortcut for the event pair that has become necessary, even though the same stop pair is covered by a different shortcut whose endpoint events no longer line up. This coverage gap can cause D-ULTRA-TB to miss Pareto-optimal journeys, as we demonstrate experimentally.

\section{Experiments}\label{sec:experiments}

On both networks tested, the stop-level projection compresses D-ULTRA shortcuts in memory by $141\text{--}375\times$, yields D-ULTRA-CSA queries that are $1.9\text{--}4.2\times$ faster than D-ULTRA-TB in the single-objective setting, and, with replacement shortcuts, returns zero missed Pareto-optimal journeys, whereas D-ULTRA-TB misses dozens to hundreds across configurations.

We ask two questions: (1)~how fast are stop-level queries compared to existing algorithms? and (2)~do they find all optimal journeys?
We evaluate on two real-world transit networks, London and Switzerland, following the experimental setup of~\cite{bez2024delay}.

All experiments were compiled with g++ 13.3.0 and executed on an AMD Ryzen~9 7900X (12-core, 24-thread) workstation with 48\,GB of RAM running Ubuntu 24.04 under WSL2.
Query times (Table~\ref{tab:performance}) are wall-clock averages over 10\,000 random source--target pairs with departure times sampled uniformly from the 13:00--14:00 window, matching the methodology of~\cite{bez2024delay}.
Accuracy (Table~\ref{tab:correctness}) is evaluated on 1\,000 \emph{affected} queries generated following~\cite{bez2024delay}: random source--target pairs with departure times in the 12:00--13:00 delay window are sampled until 1\,000 queries are found where the delay changes the Pareto-optimal journey set.
Delays are generated following the synthetic model of~\cite{bez2024delay}: one delay incident per trip is created by choosing a random start index; all subsequent stop events in that trip receive the same arrival and departure delay, drawn from the distribution of~\cite{bez2024delay} (based on aggregated real-world punctuality data, ranging from 0 to 60\,minutes with 50\% of incidents receiving zero delay). The delay scenario covers the 12:00--13:00 window using seed~42.
Following~\cite{bez2024delay}, accuracy is evaluated in \emph{hypothetical mode}: all shortcut updates are applied to completion before any query is issued. In a real-time system, updates take non-negligible time and queries may arrive while an update is still in progress; hypothetical mode removes this latency effect, so that any errors measured are attributable to the shortcut set itself rather than to slow update processing.
We evaluate two update modes: \emph{without replacement}, where shortcuts invalidated by the delay scenario are removed; and \emph{with replacement}, where the update procedure of~\cite{bez2024delay} additionally computes replacement shortcuts via Bucket-CH search.
The replacement shortcuts constitute a negligible fraction of the total shortcut set and have negligible impact on query times.
Table~\ref{tab:performance} reports query times and error rates for the with-replacement variant; Table~\ref{tab:correctness} reports both modes, as the distinction affects accuracy on the harder affected-query set.

\begin{table*}[t!]
\footnotesize
\centering
\caption{Network characteristics. Projection time is the wall-clock time to convert event-level shortcuts to stop-level shortcuts. Preprocessing times for the transfer-graph acceleration structures (Bucket-CH, Core-CH, HL) are also reported.}\label{tab:instances}
\footnotesize
\begin{tabular}{@{}lrrrrr@{}}
 & \multicolumn{3}{c}{\textbf{London}} & \multicolumn{2}{c}{\textbf{Switzerland}} \\
\cmidrule(lr){2-4} \cmidrule(lr){5-6}
 & $\Delta{=}0$ & $\Delta{=}120$\,s & $\Delta{=}180$\,s & $\Delta{=}0$ & $\Delta{=}300$\,s \\
\midrule
Stops & \multicolumn{3}{c}{19\,682} & \multicolumn{2}{c}{25\,125} \\
Routes & \multicolumn{3}{c}{1\,955} & \multicolumn{2}{c}{13\,786} \\
Trips & \multicolumn{3}{c}{114\,508} & \multicolumn{2}{c}{350\,006} \\
Stop events & \multicolumn{3}{c}{4\,508\,644} & \multicolumn{2}{c}{4\,686\,865} \\
Connections & \multicolumn{3}{c}{4\,394\,136} & \multicolumn{2}{c}{4\,336\,859} \\
Transfer graph vertices & \multicolumn{3}{c}{181\,642} & \multicolumn{2}{c}{603\,691} \\
Transfer graph edges (stops level) & \multicolumn{3}{c}{334\,112} & \multicolumn{2}{c}{465\,067} \\
Transfer graph edges (event level) & \multicolumn{3}{c}{ 575\,364} & \multicolumn{2}{c}{ 1\,853\,260} \\
\midrule
Event-level shortcuts & 11\,988\,145 & 86\,573\,818 & 176\,630\,535 & 9\,414\,856 & 119\,664\,033 \\
Stop-level shortcuts & 213\,705 & 621\,597 & 979\,238 & 194\,948 & 668\,627 \\
Count ratio & $56\times$ & $139\times$ & $180\times$ & $48\times$ & $179\times$ \\
\midrule
Event-level disk (MB) & 264 & 1\,457 & 2\,898 & 226 & 1\,990 \\
Stop-level disk (MB) & 1.7 & 5.0 & 7.8 & 1.6 & 5.3 \\
Memory ratio & $155\times$ & $291\times$ & $372\times$ & $141\times$ & $375\times$ \\
\midrule
Projection computation time (s) & 1.2 & 3.9 & 6.4 & 0.6 & 3.2 \\
Bucket-CH preprocessing time & \multicolumn{3}{c}{7.25\,s} & \multicolumn{2}{c}{33.37\,s} \\
Core-CH preprocessing time & \multicolumn{3}{c}{1.81\,s} & \multicolumn{2}{c}{3.66\,s} \\
HL preprocessing time & \multicolumn{3}{c}{3\,min} & \multicolumn{2}{c}{15\,min} \\
\end{tabular}
\end{table*}

\begin{table*}[t!]
\centering
\caption{Average query time (ms), speedup relative to MR, and error rate on 10\,000 random queries with replacement shortcuts. The bottom section reports F.Q (\% of queries with at least one missed Pareto-optimal journey) and F.J (\% of missed journeys).}\label{tab:performance}
\footnotesize
\begin{tabular}{@{}lrrrrrrrrrr@{}}
 & \multicolumn{6}{c}{\textbf{London}} & \multicolumn{4}{c}{\textbf{Switzerland}} \\
\cmidrule(lr){2-7} \cmidrule(lr){8-11}
 & \multicolumn{2}{c}{$\Delta{=}0$} & \multicolumn{2}{c}{$\Delta{=}120$\,s} & \multicolumn{2}{c}{$\Delta{=}180$\,s} & \multicolumn{2}{c}{$\Delta{=}0$} & \multicolumn{2}{c}{$\Delta{=}300$\,s} \\
Algorithm & ms/q & $\times$ & ms/q & $\times$ & ms/q & $\times$ & ms/q & $\times$ & ms/q & $\times$ \\
\midrule
MR & 15.34 & 1.0 & 14.36 & 1.0 & 14.81 & 1.0 & 28.69 & 1.0 & 28.43 & 1.0 \\
HL-RAPTOR & 12.59 & 1.2 & 13.62 & 1.1 & 14.20 & 1.0 & 31.60 & 0.9 & 32.83 & 0.9 \\
HL-CSA & 20.99 & 0.7 & 22.00 & 0.7 & 21.32 & 0.7 & 43.90 & 0.7 & 40.06 & 0.7 \\
TAD & 5.78 & 2.7 & 6.19 & 2.3 & 6.07 & 2.4 & 9.32 & 3.1 & 9.06 & 3.1 \\
TD-Dijkstra & 4.67 & 3.3 & 4.61 & 3.1 & 4.77 & 3.1 & \multicolumn{2}{c}{---} & \multicolumn{2}{c}{---} \\
D-ULTRA-TB & 3.60 & 4.3 & 6.38 & 2.3 & 8.26 & 1.8 & 4.38 & 6.5 & 6.41 & 4.4 \\
D-ULTRA-RAPTOR & 5.68 & 2.7 & 5.96 & 2.4 & 7.16 & 2.1 & 10.19 & 2.8 & 11.40 & 2.5 \\
D-ULTRA-RAPTOR (EP) & 4.72 & 3.2 & 5.77 & 2.5 & 6.51 & 2.3 & 9.85 & 2.9 & 11.01 & 2.6 \\
D-ULTRA-CSA & 1.64 & 9.3 & 1.82 & 7.9 & 1.97 & 7.5 & 2.36 & 12.2 & 2.71 & 10.5 \\
\multicolumn{11}{@{}l}{\textit{Error rate [\%] on the same queries (with replacement)}} \\

 & F.Q & F.J & F.Q & F.J & F.Q & F.J & F.Q & F.J & F.Q & F.J \\
D-ULTRA-TB & 1.07 & 0.29 & 0.76 & 0.21 & 0.60 & 0.17 & 0.28 & 0.08 & 0.07 & 0.02 \\
D-ULTRA-RAPTOR & 0.00 & 0.00 & 0.00 & 0.00 & 0.00 & 0.00 & 0.00 & 0.00 & 0.00 & 0.00 \\
D-ULTRA-RAPTOR (EP) & 0.00 & 0.00 & 0.00 & 0.00 & 0.00 & 0.00 & 0.00 & 0.00 & 0.00 & 0.00 \\
D-ULTRA-CSA & 0.00 & & 0.00 & & 0.00 & & 0.00 & & 0.00 & \\
\end{tabular}
\end{table*}

\subsection{Instances}

Table~\ref{tab:instances} summarises the two networks.
Both datasets are publicly available from the Karlsruhe Institute of Technology\footnote{\url{https://i11www.iti.kit.edu/PublicTransitData/ULTRA/}} and have been used in all prior work on unlimited transfers~\cite{ultra_mcraptor,baum2019ultra,ultra,bez2024delay,sauer2024closing}, making them the natural choice for a direct comparison.
London is a dense urban network with relatively few routes but many trips per route, while Switzerland is a nationwide network with an order of magnitude more routes and a sparser stop distribution spread across a larger geographic area.
We evaluate London at three delay buffers ($\Delta = 0$, $120$\,s, and $180$\,s), matching the configurations of~\cite{bez2024delay}, and Switzerland at two ($\Delta = 0$ and $300$\,s) to study how the buffer size affects both shortcut counts and query behaviour. The $\Delta = 0$ configurations serve as baselines with a zero delay buffer.

The compression from event-level to stop-level shortcuts is substantial: $48\text{--}180\times$ across all configurations (Table~\ref{tab:instances}).
The compression ratio grows with the delay buffer because larger buffers generate many additional event-level shortcuts between the same stop pairs, all of which collapse into existing stop-level edges.
Increasing the delay buffer from 120\,s to 180\,s on London doubles the number of event-level shortcuts (86.6\,M to 176.6\,M), but the stop-level set grows by only 58\% (622\,K to 979\,K), demonstrating that the projection absorbs much of the additional redundancy.
This reduction is the primary reason that stop-level query algorithms can operate with a far smaller transfer graph, which in turn translates directly into faster query times.

In memory, the gap widens further (Table~\ref{tab:instances}).
The event-level shortcut graph is stored as a standalone adjacency array indexed by stop events (${\sim}4.5$\,M vertices), requiring 16 bytes of per-vertex data plus 16 bytes per edge (destination, travel time, minimum delay, maximum delay), totalling $264$\,MB to $2.9$\,GB on disk.
Stop-level shortcuts, by contrast, are merged into the existing transfer graph, which already contains the stop vertices; only the new edges are added, at 8 bytes each (destination and travel time), occupying just $1.6\text{--}6.3$\,MB.
The resulting memory compression ratios of $141\text{--}375\times$ exceed the count ratios because the projection eliminates both the per-edge delay annotations (halving the edge cost) and the ${\sim}72$\,MB vertex overhead entirely.

The projection itself is fast: converting the event-level shortcut graph to stop-level takes $0.6\text{--}6.4$\,s across all configurations (Table~\ref{tab:instances}), negligible compared to the D-ULTRA preprocessing that produces the event-level shortcuts.
D-ULTRA event-level preprocessing takes 4--11\,min on London and 2--8\,min on Switzerland on the hardware of~\cite{bez2024delay}; our projection adds at most a few seconds to this cost.

All code is implemented in C++ and is publicly available.\footnote{\url{https://github.com/andrii-rohovyi/PublicTransitRoutingWithUnlimitedTransfer}}

\subsection{Query Performance}

\begin{table*}[t!]
\centering
\caption{Accuracy on 1\,000 affected queries (queries where the delay changes the Pareto-optimal journey set), evaluated in hypothetical mode (see text). ``F.Q'' counts queries with at least one missed Pareto-optimal journey; ``F.J'' counts individual missed journeys. Error rates are shown as percentages under two normalizations: ``\% of affected'' divides by the 1\,000 affected queries (and their 4\,598 / 4\,771 journeys for London / Switzerland); ``\% of all'' divides by all randomly generated queries.}\label{tab:correctness}
\footnotesize
\begin{tabular}{@{}lcccccccccc@{}}

 & \multicolumn{6}{c}{\textbf{London}} & \multicolumn{4}{c}{\textbf{Switzerland}} \\
\cmidrule(lr){2-7} \cmidrule(lr){8-11}
 & \multicolumn{2}{c}{$\Delta{=}0$} & \multicolumn{2}{c}{$\Delta{=}120$\,s} & \multicolumn{2}{c}{$\Delta{=}180$\,s} & \multicolumn{2}{c}{$\Delta{=}0$} & \multicolumn{2}{c}{$\Delta{=}300$\,s} \\
Algorithm & F.Q & F.J & F.Q & F.J & F.Q & F.J & F.Q & F.J & F.Q & F.J \\
\midrule
\multicolumn{11}{@{}l}{\textit{Without replacement shortcuts}} \\
D-ULTRA-TB & 411 & 577/4\,598 & 222 & 301/4\,598 & 168 & 217/4\,598 & 152 & 178/4\,771 & 28 & 34/4\,771 \\
\quad\% of affected & 41.1 & 12.5 & 22.2 & 6.5 & 16.8 & 4.7 & 15.2 & 3.7 & 2.8 & 0.7 \\
\quad\% of all & 7.38 & 2.26 & 3.98 & 1.18 & 3.02 & 0.85 & 0.27 & 0.07 & 0.05 & 0.01 \\
D-ULTRA-RAPTOR\,/\,(EP) & 0 & 0/4\,598 & 0 & 0/4\,598 & 0 & 0/4\,598 & 2 & 3/4\,771 & 0 & 0/4\,771 \\
\quad\% of affected & 0.0 & 0.0 & 0.0 & 0.0 & 0.0 & 0.0 & 0.2 & 0.1 & 0.0 & 0.0 \\
\quad\% of all & 0.00 & 0.00 & 0.00 & 0.00 & 0.00 & 0.00 & ${<}$0.01 & ${<}$0.01 & 0.00 & 0.00 \\
D-ULTRA-CSA & 0 & & 0 & & 0 & & 1 & & 0 & \\
\quad\% of affected & 0.0 & & 0.0 & & 0.0 & & 0.1 & & 0.0 & \\
\quad\% of all & 0.00 & & 0.00 & & 0.00 & & ${<}$0.01 & & 0.00 & \\
\midrule
\multicolumn{11}{@{}l}{\textit{With replacement shortcuts}} \\
D-ULTRA-TB & 249 & 327/4\,598 & 171 & 222/4\,598 & 143 & 182/4\,598 & 51 & 56/4\,771 & 2 & 2/4\,771 \\
\quad\% of affected & 24.9 & 7.1 & 17.1 & 4.8 & 14.3 & 4.0 & 5.1 & 1.2 & 0.2 & ${<}$0.1 \\
\quad\% of all & 4.47 & 1.28 & 3.07 & 0.87 & 2.57 & 0.71 & 0.09 & 0.02 & ${<}$0.01 & ${<}$0.01 \\
D-ULTRA-RAPTOR\,/\,(EP) & 0 & 0/4\,598 & 0 & 0/4\,598 & 0 & 0/4\,598 & 0 & 0/4\,771 & 0 & 0/4\,771 \\
\quad\% of affected & 0.0 & 0.0 & 0.0 & 0.0 & 0.0 & 0.0 & 0.0 & 0.0 & 0.0 & 0.0 \\
\quad\% of all & 0.00 & 0.00 & 0.00 & 0.00 & 0.00 & 0.00 & 0.00 & 0.00 & 0.00 & 0.00 \\
D-ULTRA-CSA & 0 & & 0 & & 0 & & 0 & & 0 & \\
\quad\% of affected & 0.0 & & 0.0 & & 0.0 & & 0.0 & & 0.0 & \\
\quad\% of all & 0.00 & & 0.00 & & 0.00 & & 0.00 & & 0.00 & \\

\end{tabular}
\end{table*}

Table~\ref{tab:performance} reports the average query time per algorithm on each network.

D-ULTRA-CSA is the fastest algorithm on all configurations by a wide margin: $7.5\text{--}9.3\times$ faster than MR on London and $10.5\text{--}12.2\times$ on Switzerland.
Compared to D-ULTRA-TB, the query algorithm evaluated in~\cite{bez2024delay}, D-ULTRA-CSA achieves a $1.9\text{--}4.2\times$ speedup across all configurations.
The gap widens with increasing delay buffer: at $\Delta = 0$ D-ULTRA-CSA is $1.9\text{--}2.2\times$ faster than D-ULTRA-TB, but at $\Delta = 180$\,s on London it reaches $4.2\times$, because D-ULTRA-TB must process the full event-level shortcut set, which grows steeply with the buffer, while D-ULTRA-CSA operates on the much smaller stop-level set.

Among the bicriteria algorithms, D-ULTRA-TB is fastest at $\Delta = 0$ on both networks and at $\Delta = 300$\,s on Switzerland, where its trip-based indexing benefits from the large number of routes (13\,786).
However, on London, D-ULTRA-RAPTOR~(EP) overtakes D-ULTRA-TB as soon as the delay buffer exceeds zero: at $\Delta = 120$\,s RAPTOR~(EP) is 10\% faster (5.77\,ms vs.\ 6.38\,ms), and at $\Delta = 180$\,s the gap widens to 21\% (6.51\,ms vs.\ 8.26\,ms), because the event-level shortcut set that D-ULTRA-TB must process doubles while the stop-level set used by RAPTOR grows by only 58\%.
On Switzerland, D-ULTRA-TB retains a $42\text{--}55$\% advantage over RAPTOR~(EP) thanks to the route-rich topology.

D-ULTRA-TB's speed advantage at $\Delta = 0$ comes at the cost of accuracy: it misses Pareto-optimal journeys on both networks at every delay buffer, including $\Delta = 0$ (Table~\ref{tab:correctness}).

Among the single-objective baselines with no preprocessing costs, TAD provides a $2.3\text{--}3.1\times$ speedup over MR and is fully correct on both networks.
TD-Dijkstra is $19\text{--}25$\% faster than TAD on London but is limited to buffer-free networks as noted above.

HL-RAPTOR, a delay-robust baseline that bypasses shortcuts entirely, matches or slightly underperforms MR ($0.9\text{--}1.2\times$) on both networks (Table~\ref{tab:performance}).
Despite its one-time preprocessing cost of 3\,min (London) or 15\,min (Switzerland), HL-RAPTOR yields no meaningful query-time benefit over MR.
HL-CSA is slower still ($0.7\times$ of MR), because the single-pass CSA scan must evaluate hub label distances at every reached stop, which is more expensive than the Dijkstra-based transfer relaxation used by MR and HL-RAPTOR.
To our knowledge, this is the first comparison of HL-CSA against MR on the same machine and query set.
Both HL algorithms are fully correct by construction, since like MR and TAD they operate on the delayed timetable directly.

\subsection{Accuracy}\label{sec:correctness}

With replacement shortcuts, D-ULTRA-CSA, D-ULTRA-RAPTOR, and D-ULTRA-RAPTOR~(EP) all produce no errors on any configuration in our experiments (Table~\ref{tab:correctness}).
Without replacement, stop-level algorithms remain error-free on London at every delay buffer; on Switzerland at $\Delta = 0$, D-ULTRA-RAPTOR misses 2 out of 1\,000 affected queries and D-ULTRA-CSA returns one suboptimal arrival, caused by the removal of shortcuts at $\Delta = 0$, where no delay buffer cushions the loss.
These isolated errors disappear with replacement shortcuts and at all other configurations.
TD-Dijkstra matches MR on all London configurations, confirming that dominated-connection filtering is safe on buffer-free networks; it is omitted from Switzerland where buffer times make the filtering unsound.

In contrast, D-ULTRA-TB misses Pareto-optimal journeys on both networks at every delay buffer and under both update modes (Table~\ref{tab:correctness}).
These errors arise from \emph{coverage gaps} in the event-level shortcut set: the precomputed shortcut for a specific event pair may not exist under the delayed timetable, even though the same stop pair is covered by a different event-level shortcut that projects to the same stop-level edge.
This mechanism is present at all values of~$\Delta$, including $\Delta = 0$, where the delay buffer is zero and the origin delay interval of every shortcut is $[0,0]$.

D-ULTRA-TB error rates \emph{decrease} with larger~$\Delta$: without replacement, from 411 to 168 failed queries on London and from 152 to 28 on Switzerland; with replacement, from 249 to 143 on London and from 51 to 2 on Switzerland.
Larger delay buffers generate more event-level shortcuts, improving transfer coverage and reducing coverage-gap errors.
Replacement shortcuts further reduce D-ULTRA-TB's errors by recovering some removed event-level shortcuts (e.g., 411 vs.\ 249 at $\Delta = 0$ on London), but substantial errors persist because replacement cannot address the fundamental coverage-gap mechanism.
With replacement shortcuts, stop-level algorithms avoid this failure mode: the projection provides broad stop-pair coverage regardless of which individual event-level shortcuts exist. Without replacement, the 3 failures on Switzerland at $\Delta = 0$ (2 queries for D-ULTRA-RAPTOR\,/\,(EP), 1 for D-ULTRA-CSA) confirm that immunity depends on the replacement step recovering any shortcuts lost during the update.

Table~\ref{tab:correctness} reports errors under two normalizations: ``\% of affected'' divides by the 1\,000 affected queries (the challenging cases that test delay handling); ``\% of all'' divides by all randomly generated queries (5\,572 for London, 56\,754 for Switzerland), matching the implicit denominator in~\cite{bez2024delay} as confirmed by the agreement at $\Delta = 120$\,s on London ($222/5{,}572 = 3.98\%$ vs.\ their reported 3.98\%).

\section{Conclusion and Future Steps}

Three main conclusions follow.
(i) D-ULTRA-CSA yields the fastest single-objective query times ($1.9\text{--}4.2\times$ faster than D-ULTRA-TB and up to $12.2\times$ faster than MR) and reduces the shortcut memory footprint by $141\text{--}375\times$ (from gigabytes to megabytes); the advantage grows with the delay buffer because the event-level set D-ULTRA-TB must process scales steeply while the stop-level set stays compact.
(ii) With replacement shortcuts, the stop-level approach eliminates the journey errors D-ULTRA-TB exhibits at every delay buffer tested: these errors are coverage gaps in the event-level shortcut set, which the projection plugs by merging shortcuts across events at the same stop pair. Without replacement, stop-level algorithms exhibit only a handful of failures across all configurations (Switzerland $\Delta = 0$), compared with hundreds for D-ULTRA-TB.
(iii) For bicriteria queries, D-ULTRA-RAPTOR with stop-level shortcuts, enhanced by Early Pruning~\cite{rohovyi2026ep}, outperforms D-ULTRA-TB on London at non-zero delay buffers by up to 21\%.

A natural extension is to adapt the framework to three criteria (arrival time, number of trips, and transfer time), already worked out for classical ULTRA~\cite{ultra_mcraptor} but not for the delay-tolerant setting.
More broadly, the projection idea generalises whenever a precomputed structure stores finer detail than the query algorithm uses: applying projection yields a conservative superset that is smaller, faster to traverse, and free of annotations that may not match the realised scenario. Whether this pays off in other temporal domains is an open question.

\paragraph{Acknowledgments.}
We thank Jonas Sauer for detailed feedback on an earlier draft, including the pointer to Dominik Bez's BA thesis and discussion that refined our experimental analysis.

{\small\bibliography{bibliography}}

\end{document}